\documentstyle[12pt]{article}

\def\etal{{\it et al.}}
\def\lala{\Lambda\Lambda}
\begin{document}
\bibliographystyle{unsrt}
%
\begin{center}
{\Large \bf Widths of $\Xi$ Hypernuclear States}\\
\vspace{0.4cm}
{D. J. Millener$^{1}$, C. B. Dover$^{1}$ and A. Gal$^{2}$}\\
\bigskip
$1)$ Brookhaven National Laboratory, Upton, NY 11973 \\
$2)$ Racah Institute of Physics, The Hebrew University, Jerusalem 91904,
Israel\\
\end{center}
\begin{abstract}
 The $\Lambda$ and neutron decay widths of $\Xi$ hypernuclear states,
based on calculated $\Xi N \to \lala$ mixing amplitudes, are estimated.
The widths which result from using the Nijmegen Model D interaction
are sufficiently small, of order 1.5 MeV, that experiments to observe
$\Xi$ hypernuclear states using the $(K^-,K^+)$ reaction may be
feasible.
\end{abstract}
\begin{center}
{\it We would like to dedicate this paper to the memory of Professor
Hiroharu Band\=o, a warm and courageous man, a friend, and a major
contributor to hypernuclear physics.}
\end{center}
\section{Introduction}
\label{Introduction}

 The structure properties of hypernuclei reflect the nature of the underlying
baryon-baryon interactions, for which there is very little two-body data
other than for the nucleon-nucleon interaction, and thus can provide rather
stringent tests of models for the free-space $YN$ and $YY$ interactions.
Over the years, the Nijmegen group has constructed a number of
one-boson-exchange models for the baryon-baryon interaction ($NN$,
$\Lambda N - \Sigma N$, and $\Xi N -\Lambda\Lambda - \Sigma\Sigma
-\Lambda\Sigma$) using SU(3)$_{flavor}$ symmetry to relate baryon-baryon-meson
coupling constants and phenomenological hard or soft cores at short
distances. More recently, the J\"ulich group has constructed meson-exchange
models for the $YN$ interaction along the lines of the Bonn model for the
$NN$ interaction using SU(6) symmetry to relate coupling constants and
short-range form factors (reviews of all the relevant models appear
in the proceedings of the Shimoda conference on hypernuclear and strange
particle physics \cite{shim92}; in particular, see the reviews by Rijken
and Holinde).

     To test these models against the considerable body of information
on $\Lambda$ hypernuclei \cite{davp86,chr89,bmz90}, effective
interactions appropriate for use in limited spaces of shell-model orbits
must be calculated. This has been done by calculating nuclear-matter G-matrices
\cite{yam92,reub94} for the coupled $\Lambda N$ and $\Sigma N$ channels.
For most models, the resulting $\Lambda$ well depth is in reasonable
agreement with the empirical value deduced from fitting the
binding energies of $\Lambda$ single-particle states \cite{mgd88,yam88}
observed via the $(\pi^+$,K$^+)$ reaction \cite{pile91}, but the partial-wave
contributions vary widely for the different models. In fact, the
spin-dependence
of the central interaction varies considerably in the different models
\cite{yam92}. While empirical $\Lambda N$ effective interactions fitted to the
properties of light $\Lambda$ hypernuclei are more attractive in the singlet
than the triplet state \cite{mgdd85,fet91}, a conclusive choice between
the models cannot be made since the effective $\Lambda NN$ interaction
resulting from $\Lambda N -\Sigma N$ coupling, which contributes strongly
to the spin-dependence for A=4 \cite{gib88}, has yet to be evaluated in
a consistent way.

  The baryon-baryon interactions which result from an extension of the
boson-exchange models to the $S\!=-2$ sector can be qualitatively quite
different from one another at long range \cite{dg84,dmgd91,yam91},
even before the uncertainties at short range, where quark model
calculations may provide some guidance (K. Shimizu in Ref. \cite{shim92}),
are considered.
Rather little is known about $S\!=-2$ systems.  There are only a
a few examples of $\lala$ hypernuclear events seen in emulsion
experiments with $K^-$ beams, and a few others attributed to the formation
of $\Xi$ hypernuclei (see Ref. \cite{dgm94} for references). The
binding energies of the three examples of light $\lala$ hypernuclei
indicate that the effective $\lala$ interaction is quite strongly
attractive, with a relative $0s$ matrix element of magnitude $\sim 4.5$ MeV.
Also, when taken at face value, the $\Xi$
hypernuclear events indicate an attractive $\Xi$-nucleus potential with a
strength roughly comparable to that for $\Lambda$ hypernuclei \cite{dg83}.
When the short-range behavior of the $YY$ interaction is chosen to
be similar to that of the $NN$ interaction, only Model D of the
Nijmegen interactions can provide sufficient attraction to
account for the empirical $\lala$ matrix element \cite{yam91,him93}.
Model D also gives an attractive $\Xi$-nucleus potential \cite{dg84}.
In contrast, Model F gives repulsion in both cases.

  Improved data for the $S\!=-2$ sector would put strong constraints
on baryon-baryon interaction models. To this end, an active
experimental program is being pursued at the KEK laboratory in Japan
using the double strangeness exchange reaction on nuclear
targets \cite{shim92,aoki91,iij92}. At the Brookhaven AGS, a more intense
separated 1--2 GeV/c $K^-$ beam \cite{pile92} is now available for $(K^-,K^+)$
studies. The first use of this beam line has been in a search for the $H$
dibaryon but new proposals are focussed on more general studies of $\!S=-2$
spectroscopy. Several of these concentrate on the detection of $\lala$
hypernuclei produced by $\Xi^-$ interactions with nuclei. Another attractive
possibility would be to look for $\Xi$ hypernuclear states using the
$(K^-,K^+)$ reaction in analogy to studies of $\Lambda$ hypernuclei using the
$(\pi^+,K^+)$ reaction. Unfortunately, the $(K^-,K^+)$ cross sections leading
to single-particle $\Xi$ configurations are small, of the order of a few
hundred nb/sr for light systems \cite{dg83,dgm94,ikeda94}, and somewhat below
current detection capabilities. Nevertheless, an important question, should
this type of experiment become feasible, concerns the width of $\Xi$
hypernuclear states which arises from the mixing of $\Xi$ and $\lala$
hypernuclear states
and depends directly on the strength of the $\Xi N\to\lala$ interaction.
There have been two recent estimates of these widths \cite{dgm94,ikeda94}
using rather different approaches. Our own estimates \cite{dgm94} of the
decay (escape) widths arising from strong $n$, $p$ or $\Lambda$ emission
from $\Xi$-nucleus configurations were made using standard shell-model
techniques. For pedagogical reasons related to the simplicity of
nuclear parentages, we concentrated on calculations for a $^{15}$N
target. In this paper, we make estimates for the escape widths of bound
$\Xi^-$ hypernuclear states likely to be produced in the more
practical $^{12}$C$(K^-,K^+)^{12}_{\Xi}$Be reaction considered by Ikeda
{\it et al.} \cite{ikeda94}. While the two calculations give comparable
estimates of the widths for $\Xi N\to\lala$ interactions of the same strength,
the partial decay widths from different sources are quite different.
Consequently, we point out some of the reasons for these differences.


\section{Single-particle $\Xi$ States}
\label{singlepart}

 The early emulsion data which suggested the existence of bound $\Xi$
states in various nuclear systems were analyzed by Dover and Gal
\cite{dg83} in terms of a $\Xi$-nucleus single-particle Woods-Saxon
well with a depth $V_\Xi = 24$ MeV, radius parameter $r_0 = 1.1$ fm
and diffuseness $a =0.65$ fm (or a depth of 21 MeV for $r_0 =1.25$ fm).
This value of $V_\Xi$, comparable to that for the $\Lambda$-nucleus
potential \cite{mgd88,yam88}, binds a $\Xi^-$ in both $0s$ and $0p$
states to a $^{11}$B core with binding energies of 10.7 and 1.5 MeV,
respectively \cite{ikeda94}. Since stable targets have
$N\ge Z$, the $\Xi$ states of interest will be essentially pure
$\Xi^-$ charge states (with $T = -T_z$).

 The spectrum for the
$^{12}$C$(K^-,K^+)^{12}_{\Xi}$Be reaction should be qualitatively
very similar (see Fig.~1 of Ref. \cite{ikeda94}) to that for the well-studied
$^{12}$C$(\pi^+,K^+)^{12}_{\Lambda}$C reaction \cite{pile91,mil85,has94}.
The cross sections will, however, be considerably smaller on
account of the smaller elementary cross section and the somewhat
higher momentum transfer (around 420 MeV/c to 500 MeV/c over the
angular range $0^\circ - 10^\circ$ compared with $330-365$ MeV/c
over the same range, respectively). The distorted-wave impulse approximation
calculations for $p_K = 1.6$ GeV/c by Ikeda {\it et al.} \cite{ikeda94}
for the natural parity $j_N^{-1}j_{\Xi}$ configurations give
$0^\circ$ cross sections of 0.25 $\mu$b/sr for the $1^-$ ground-state
peak and 0.50 $\mu$b/sr for the $0^+,2^+$ complex of $p_\Xi$ states,
if the recent KEK measurement of the elementary cross section
(averaged over $\theta_{lab}$ from $1.7^\circ$ to $13.6^\circ$)
by Iijima {\it et al.} \cite{iij92} is used. These cross sections
at $p_K=1.6$ GeV should be multiplied by $\sim 0.71$, representing the
parentage for proton pickup from $^{12}$C to the ground state
of $^{11}$B.  However, this factor is gained back at $p_K=1.8$ GeV/c
where the elementary cross section appears to peak (see Fig.~5 in
Ref. \cite{dg83}) and for which the D6 beamline at Brookhaven
is optimized \cite{pile92}.

  The relevant basis for comparison for these predicted cross sections
is the sensitivity of $\sim 1.0$ $\mu$b/sr achieved for the
$(\pi^+,K^+)$ reaction with a flux of $10^7$ $\pi^+$ on target per spill
and an energy resolution of $\sim 3$ MeV \cite{pile91}. For $K^-$ at
1.8 GeV, a flux of $2.8\times 10^6$ per $10^{13}$ protons with a
$\pi^- /K^-$ of 2.5:1 was typical in 1992 (note added in proof
to Ref. \cite{pile92}); the flux is limited by the proton flux that the
production target can stand but runs have been made using
$1.5\times 10^{13}$ protons per spill. The question of how much improved
resolution can improve the sensitivity depends on the natural width
of the $\Xi$ hypernuclear states, each of which is broadened by the
strong decay processes
\begin{eqnarray}
\Xi^- p\to \lala & Q=28.3\ {\rm MeV}, \label{eq:1a} \\
\Xi^0 n\to \lala & Q=23.2\ {\rm MeV}. \label{eq:1b}
\end{eqnarray}
In fact, the first concern is that $\Xi$ hypernuclear states could
have widths that are some appreciable fraction of the nuclear matter
estimate ($\sim 10$ MeV \cite{dg83,dgm94}) and therefore comparable
with the spacings between $\Xi$ single-particle levels. Our recent
estimates \cite{dgm94} and those of Ikeda {\it et al.} \cite{ikeda94},
both of which use $\Xi N\to \lala$ conversion potentials based
on the Nijmegen Model D interaction, give much smaller widths of
around 1.5 MeV. Since the calculated widths depend quadratically on
the strength of the conversion potential, which involves the exchange
of strange mesons and varies widely in different
baryon-baryon models, the widths could differ considerably from these
estimates. Of course, both the positions and widths of $\Xi$
hypernuclear levels, if they exist and can be observed with
sufficiently high resolution, together with apparently strong
attraction in the $\lala$ interaction, would put very strong
constraints on baryon-baryon interaction models.

 In the next section, we present in some detail our calculation of
the widths of $\Xi$ hypernuclear states based on the mixing of
$\Xi$ and $\lala$ hypernuclear states through the $\Xi N\to \lala$
interaction. We concentrate on the case of states in $^{12}_{\Xi}$Be
since $^{12}$C is the most likely target for attempts to observe
bound $\Xi$ hypernuclear states using the $(K^-,K^+)$ reaction.
We also compare our calculation with that of Ikeda {\it et al.}
\cite{ikeda94}.

\section{Mixing of $\Xi$ and $\Lambda\Lambda$ Hypernuclear States}
\label{mixing}

  It is the admixtures of $\lala$ hypernuclear states into a $\Xi$
hypernuclear state which give the state its decay width into single
$\Lambda$, nucleon and other
channels. The advantage of a microscopic shell-model treatment is that
the decay energies are defined in terms of the energies of the nuclear
cores and the interaction of the two $\Lambda$'s with these cores and
each other. The decay energies are then quite well determined in
terms of our empirical knowledge of single and double $\Lambda$
hypernuclei. In practice, we make a weak-coupling approximation
in which we couple the hyperons to specific nuclear core states and
do not consider the mixing of these states in a shell-model
diagonalization; there are, however, constraints imposed the fact that
there should be no spurious center of mass motion, which we take into account
as necessary. In this basis, an estimate of the mixing matrix elements
in terms of the two-body matrix elements of the $\Xi N\to \Lambda\Lambda$
transition potential is straightforward. The actual admixures are
calculated in first-order perturbation theory and the spectroscopic
amplitudes for each decay channel are then calculated by standard
parentage and recoupling techniques.

\subsection{Energies and Decay Thresholds}
\label{energies}

 The energies of the most important $\Xi^-$ and $\lala$ weak-coupling
basis states and the relevant decay channels are given in Fig. 1 for
the specific case of a $^{12}$C target, with the energies measured with
respect to the $^{12}_{\lala}$Be ground state which
is bound with respect to the $\Lambda +\Lambda$ threshold by
B$_{\Lambda\Lambda}$ where
\begin{equation}
{\rm B}_{\Lambda\Lambda} = 2{\rm B}_{\Lambda}(^{11}_{\ \Lambda}{\rm Be})
 +\Delta {\rm B}_{\Lambda\Lambda}\ .
\label{eq:bll}
\end{equation}
The single-$\Lambda$ binding energies are known quite accurately for many
$p$-shell hypernuclei \cite{davp86} and can be quite reliably estimated
in cases for which a measurement does not exist. For example, in the present
case, one obtains B$_\Lambda(^{11}_{\ \Lambda}$Be) = 10.46 MeV by adding
the 0.22 MeV charge-dependent difference between
B$_\Lambda(^{10}_{\ \Lambda}$Be) and B$_\Lambda(^{10}_{\ \Lambda}$B) to
B$_\Lambda(^{11}_{\ \Lambda}$B). Alternatively, charge-independent
estimates using empirically determined $\Lambda$-nucleus potentials
\cite{mgd88,yam88} would be quite accurate enough for the present
purposes. $\Delta{\rm B}_{\Lambda\Lambda}$
is the interaction energy of a pair of $\Lambda$ particles in
the $^1S_0$ state and is around 4.5 MeV for the three known
examples of ${\Lambda\Lambda}$ hypernuclei \cite{dmgd91,dal89}.
The $^{11}{\rm B}+\Xi^-$ threshold is a further 28.34 -
S$_p(^{11}{\rm B}) = 17.1$ MeV above the $^{10}{\rm Be}+\Lambda +\Lambda$
threshold.

  As noted in the previous section, the binding energy of the
$0s$ ($0p$) $\Xi^-$ state is about 10.7 (1.5) MeV for the $\Xi^-$-nucleus
potential of Dover and Gal \cite{dg83}, which has a Woods-Saxon form
with a well depth of 24 MeV for $r_0 = 1.1$ fm and $a=0.65$ fm.
In the weak-coupling limit, the $(K^-,K^+)$ reaction will populate
the $\Xi^-$ states in proportion to the proton pickup spectroscopic
factor relating the target and core states \cite{auer83}. For no
spin-flip, the $0s$ states will have spin-parity $1^-$ and two $0p$
$2^+$ states will be most strongly produced, with equal cross
sections in the weak-coupling limit. In practice, the $2^+$ states
will be mixed and separated by about an MeV \cite{auer83}, depending on
the $\Xi N$ interaction strength and the $\Xi$ spin-orbit splitting
\cite{dg84}, which is of the same order as the expected widths of the
states.

 The ${\Lambda\Lambda}$ admixtures into the $\Xi$ state, by
conversion on a $0s$- or $0p$-shell proton, give rise to
decay amplitudes into the single-baryon escape channels shown
at the left of Fig. 1. The lowest threshold is for neutron
emission, as will generally be the case since the transformation of two
protons from the $p$-shell target nucleus leaves a neutron-rich
A-2 core for the ${\Lambda\Lambda}$ hypernucleus. For reference,
the separation energies with respect to the $\Lambda\Lambda$
hypernuclear ground state are (assuming a constant
$\Delta{\rm B}_{\Lambda\Lambda}$ for all species)
\begin{eqnarray}
{\rm S}_\Lambda & = & {\rm B}_\Lambda (^{11}_{\ \Lambda}{\rm Be})
+ \Delta{\rm B}_{\Lambda\Lambda} \label{eq:blam} \\
{\rm S}_n & = & 2{\rm B}_\Lambda (^{11}_{\ \Lambda}{\rm Be}) -
2{\rm B}_\Lambda (^{10}_{\ \Lambda}{\rm Be}) +{\rm S}_n
(^{10}{\rm Be}) \label{eq:bn} \\
{\rm S}_p & = & 2{\rm B}_\Lambda (^{11}_{\ \Lambda}{\rm Be}) -
2{\rm B}_\Lambda (^{10}_{\ \Lambda}{\rm Li}) +{\rm S}_p
(^{10}{\rm Be}) \label{eq:bp} \\
{\rm S}_\alpha & = & 2{\rm B}_\Lambda (^{11}_{\ \Lambda}{\rm Be}) -
2{\rm B}_\Lambda (^{7}_{\Lambda}{\rm He}) +{\rm S}_\alpha
(^{10}{\rm Be}) \label{eq:balpha}
\end{eqnarray}
and so on. Also important for neutron emission is the channel in
which $^{11}_{\lala}$Be is left with one of the $\Lambda$s in
a $0s$ state and the other in a $0p$ state. The separation
between the $0s$ and $0p$ $\Lambda$ single-particle states is
very close to 10 MeV, a value that we take for definiteness.
Of course, the nuclear core may be left in an excited state and
this energy is easily taken into account in the weak-coupling
approximation.

 The $\lala$ states which are most important for $\Xi^- p\to \Lambda\Lambda$
conversion from the $0s$ $\Xi^-$ hypernuclear ground state are shown at the
right in Fig.~1. In the case of $0s0p\to 0s0p$ $\Xi^- p\to \Lambda\Lambda$
conversion, the $0p$ proton parentage of the $^{11}$B ground state
is spread over four $^{10}$Be levels, namely the $0^+$ ground state and
the first three $p$-shell $2^+$ levels (identified with the 3.37 and
5.96 MeV levels and a level around 10 MeV excitation energy; the 7.54 MeV
level and possibly the 9.4 MeV level are predominantly $p^4(sd)^2$ in
nature). Also,
to a first approximation, we need to take into account conversion
only in spin-singlet states since this forces the interaction
to take place in a relative $0s$ state for the interacting baryons
(the relative $p$ state interaction associated with spin-triplet states
will be much weaker). Thus, $\Delta$B$_{\lala}$ is the same for the
$s^2_\Lambda$ and $s_\Lambda p_\Lambda$ singlet configurations and
the $^{10}$Be$(0^+_1)\otimes {^1}P_1 (s_\Lambda p_\Lambda)$ has an
excitation energy of 10 MeV, determined by the $0s-0p$ single-particle
energy difference. The energy denominators, $\Delta E_p$ in
Fig.~1, for a perturbative estimate of the admixtures then range from
$\sim 22$ to $\sim 12$ MeV.

 The $\Xi^- p\to \Lambda\Lambda$ conversion on a $0s$-shell proton is
not so clearcut to describe and make simple estimates for since the
$0s$ nucleon-hole strength occurs at a rather high
excitation energy in the core nucleus. The $0s$ proton-hole strength
appears as a broad distribution of strength, with a width of 10 MeV or
more, in $(p,2p)$ and $(e,e'p)$ reactions \cite{fm84}.  The
corresponding $0s$ neutron strength is seen in the production of $\Lambda$
hypernuclei at low momentum transfer in the $(K^-,\pi^-)$
reaction \cite{bert81}. The spreading of the $0s$ hole strength is
apparent even in $1\hbar\omega$ shell-model calculations for $p$-shell
nuclei \cite{kj79}. The basis for these calculations consists of
$s^3 p^{n-1}$ and $s^4 p^{n-3}(sd)$ configurations (with $n=8$ for $^{11}$B)
which are necessarily admixed to ensure that the center of mass is in a $0s$
oscillator state. The $s^3p^{n-1}$ strength then gets spread in the
$1\hbar\omega$ shell-model diagonalization, with the major fragments
typically spread over 10 MeV or more in excitation energy \cite{kj79}.
Each fragment acquires substantial nucleon escape widths through the
$s^4 p^{n-3}(sd)$ components in the wave function.
The centroid of the observed strength is quite well described in these
calculations and the $0s$ proton separation energy in $p$-shell nuclei can
be adequately parametrized by the relation (in MeV) S$_{0s} \sim 20 +1.88m$,
where $m$ is the number of $p$-shell nucleons, giving S$_{0s} = 33.2$ MeV for
$^{11}$B.

 The corresponding excitation energy in the $^{10}$Be core nucleus for
the $\Lambda\Lambda$ hypernuclear states is
S$_{0s} - {\rm S}_p(^{11}{\rm B})=22$ MeV. Then, the energy separation
$\Delta E_s$ (see Fig.~1) between the $^{12}_{\ \Xi}$Be $0s$ state
and the centroid of the $s_\Lambda^2$ configurations based on the
$0s$ proton hole strength in $^{11}$B is given (in MeV) by
\begin{equation}
\Delta E_s = 28.34 +{\rm B}_{\Lambda\Lambda} -{\rm S}_{0s} -{\rm B}_{\Xi^-},
\label{eq:des}
\end{equation}
leading to a rough estimate for $\Delta E_s$ of 10 MeV for A=12.
Our width estimates are obtained by concentrating the $0s$ hole strength
at its centroid energy and making a perturbative calculation of the
admixture of the $\lala$ state based on this pure $0s$-hole state.

\subsection{Mixing and Decay Amplitudes}
\label{mixdec}

 The wave function of a strangeness -2 hypernuclear state containing
the lowest-energy $\Xi$ configuration with admixtures of the configurations
considered above will be of the form (the basis
is formed from $1\hbar\omega$ excitation energy configurations with
respect to the lowest $\Lambda\Lambda$ configurations)
\begin{equation}
a|0s^4 0p^{n-1}0s_{\Xi} \rangle + b|0s^4 0p^{n-2}0s_\Lambda 0p_\Lambda
\rangle +c|(\alpha 0s^3 0p^{n-1}+
\beta 0s^4 0p^{n-3}(1s0d))0s^2_\Lambda  \rangle .
\label{eq:ximix}
\end{equation}
 The coefficient $\beta$ provides immediate
access to $s$-wave and $d$-wave neutron emission channels.
In our weak-coupling approximation to the full shell-model calculation,
we take the pure $^{11}{\rm B}(g.s.)\otimes s_\Xi$ state as the
$\Xi^-$ hypernuclear ground state and make perturbative estimates of
the amplitudes $b$ and $c$. Some constraints are imposed on the amplitude
coefficients in Eq.~\ref{eq:ximix}, particularly $\alpha$ and $\beta$,
by the condition that the center of mass of the system remain unexcited
(in practice, in a $0s$ harmonic oscillator state).
It should also be noted that the condition that the $1\hbar\omega$
$\Lambda\Lambda$ hypernuclear states be non-spurious implies a relationship
between the amplitude $b$ and certain of the amplitudes
$c$ in Eq.~\ref{eq:ximix}.

  For a target specified by A (N and Z), (A-2)$\otimes s_\Lambda p_\Lambda$
configurations are admixed
when the $0s$ $\Xi^-$ converts on a $0p$ proton, with an amplitude
proportional to a single two-body matrix element connecting $0p_N 0s_\Xi$
and $0s_\Lambda 0p_\Lambda$ states. Which $0\hbar\omega$ A-2
core states must be considered depends on the parentage distribution
of the $0p$ proton pickup strength from the A-1 core of the
$\Xi^-$ hypernuclear state.

   The (A-2)$s^{-1}_N\otimes s^2_\Lambda$ configurations are admixed
when the $0s$ $\Xi^-$ converts on a $0s$ proton, with an amplitude
proportional to a two-body matrix element connecting $0s_N 0s_\Xi$
and $0s^2_\Lambda$ states. We consider the single A-2 core state
which corresponds to a pure $0s$ hole state with respect to the
(A-1)$g.s.$ (in reality, this $0s$-hole strength is fragmented over
a considerable range of excitation energy). In effect, one orthogonalizes
$0s^30p^{n-1}$ states, with $0p^{n-1}$ parts identical to the
(A-1)$g.s.$ wave function, to spurious $1\hbar\omega$ center of mass
excitations of the A-2 nucleus. In general, there will be two possible
$J$ and $T$ values for the $0s$ hole states. The fractional reduction
in $0s$ proton occupancy, with respect to the naive shell-model value
of 2, is represented by
\begin{equation}
{\bar \alpha}^2 = 1-{(Z-3)\over 2(A-2)},\quad\quad {\bar \beta}^2
 = {(Z-3)\over 2(A-2)}.
\label{eq:hole}
\end{equation}
The quantities ${\bar \alpha}$ and ${\bar\beta}$ are closely related to,
but not exactly equal to, $\alpha$ and $\beta$ in Eq.~\ref{eq:ximix}
since they repesent averages over several $J$ and $T$ values and,
in addition, the two $s^3p^{n-1}$ configurations in Eq.~\ref{eq:ximix}
are not identical as a result of the orthogonalization process.

  To calculate the two-body matrix elements required for the mixing
calculation, we take the Gaussian conversion potential for
$p\Xi^-\to\lala$ used by Myint in Ref.~\cite{shim92} and ignore,
for simplicity, the mass differences between the interacting baryons.
Then, any two-body matrix elements can be simply expressed in terms of
the Talmi integrals
\begin{equation}
 I_p = {V_0\over {(1+2b^2/{\mu^2})^{p+3/2}}},
\label{eq:talmi}
\end{equation}
where, for Myint's potential, $V_0=57.872$ MeV and $\mu=0.855$ fm.
For $b=1.64$ fm, appropriate to $^{12}$C, $I_0 = 2.39$ MeV. For
comparison, we deduce $I_0 = 2.44$ MeV from the $0s^2$ matrix element
quoted by Ikeda {\it et al.} \cite{ikeda94}, who use a $\delta$-function
interaction, also based on the Nijmegen Model D potential.
The diagonal $0s^2$ and $0s0p$ $LS$-coupling matrix elements, for the
mixing calculations discussed above, take the values $I_0$ and
$\sqrt{1/2} I_0$, respectively, where the factor $\sqrt{1/2}$ arises
from the relative $0s$ content of the $p\Xi^-$ state. In the latter case,
the projections of $0p_{3/2}0s_{1/2}$ and $0p_{1/2}0s_{1/2}$ configurations
on L=1, S=0 are $\sqrt{2/3}$ and $\sqrt{1/3}$, respectively.

  The conversion from
$0p$ $\Xi^-$ hypernuclear states can be treated within the same
framework with the primary admixtures coming from $2\hbar\omega$
$\Lambda\Lambda$ hypernuclear states although $0\hbar\omega$
admixtures are possible with smaller matrix elements and larger
energy denominators (in a perturbative calculation).

 Having calculated spectroscopic factors $S_B$ for baryon emission channels,
we estimate partial decay widths for each exit channel using the relations
\begin{equation}
\Gamma = S_B \Gamma_{sp}\ ,\quad \Gamma_{sp} = 2 f \gamma P_l\ .
\label{eq:spw}
\end{equation}
For $f=1$, $\Gamma_{sp}$ is a single-particle width evaluated according
to Appendix 3F-2 of Bohr and Mottelson's book \cite{bm69}. The
penetrability $P_l$ (denoted by $s_l$ in Ref.~\cite{bm69}) and $\gamma$
(defined by Eq. (3F-42) of Ref.~\cite{bm69}) are evaluated for a square
well potential of radius $R$ and depth $V_0$. For $s$ waves,
\begin{equation}
\Gamma_{sp} = 2 f \ {{\hbar^2}\over {MR^2}}\ kR\ ,
\label{eq:spwlim}
\end{equation}
where $k$ is the wave number of relative motion. For the rather high
decay energies relevant to the decay of $\Xi$ hypernuclear states,
$\Gamma_{sp}$ for $p$ waves is close to this value and
$\Gamma_{sp}$ for $d$ waves is an appreciable fraction of it.
 Michaud \etal\ \cite{mic70}
have shown that the reflection properties of a diffuse well can be
approximated by using an equivalent square-well potential and a reflection
factor $f$, which typically has a value around 2.5 for nucleons.
The single-particle widths obtained using this prescription compare quite
well with those obtained by solving the Schr\=odinger equation for the complex
energies of resonances in a realistic Woods-Saxon well (this can only
be done for energies smaller than the widths), although there is an
indication that $f$ could be somewhat larger than 2.5 for the rather
diffuse light systems of interest.

 To take a purely nuclear example of the calculation of decay widths,
we consider the $0s$ proton-hole state in the $^{16}$O closed shell.
The 20\% $0s^40p^{10}(1s0d)$ admixture gives rise to nucleon decay
amplitudes to $0\hbar\omega$ states of the A=14 system with excitation
energies up to about 11 MeV. Since the $0s$ strength is centered
at $\sim 32$ MeV excitation energy in $^{15}$N, the decay energies range from
22 to 11 MeV. Using spectroscopic factors calculated from the pure $0s$
hole state to $p$-shell states of $^{14}$N and $^{14}$C and estimates
of full single-particle widths for a diffuse potential well at the
appropriate decay energies (using Eq.~\ref{eq:spw}), we calculate an
escape width for a pure $0s$-hole state located at the centroid
energy of about 8.5 MeV ($\Gamma_n = 5.4$ MeV). To compare with the
experimentally observed distribution of $0s$-hole strength, we
would have to fold escape widths with the distribution of hole strength
obtained in a shell-model calculation (spreading width) and with the
response function for the excitation process (for example, an $(e,e'p)$
reaction). Since the crude estimate for the escape width is already
less than a factor of two of the observed width, the width obtained by
taking into account the spreading width should be in considerably better
agreement with the data, which gives us confidence that our
procedure will give reasonable results in hypernuclear applications.

\subsubsection{$0s_\Lambda 0p_\Lambda$ admixtures into the $0s$ $\Xi^-$ state}
\label{spmix}

 We write the wave function of the $1^-$, $0s$ $\Xi^-$ state with
$0s_\Lambda 0p_\Lambda$ admixtures as (cf. Eq.~\ref{eq:ximix})
\begin{equation}
|0s\ \Xi^-\rangle = |^{11}{\rm B}\otimes s_\Xi\rangle +\sum_n
b(\alpha_n J_n) |^{10}{\rm Be}(J_n^\pi)\otimes s_\Lambda p_\Lambda\rangle ,
\label{eq:sxi}
\end{equation}
where
\begin{equation}
  b(\alpha_n J_n) = M(\alpha_n J_n)/\Delta E(\alpha_n J_n)
\label{eq:amp}
\end{equation}
is the perturbative admixture of each $\lala$ state and $M(\alpha_n J_n)$
is the corresponding mixing matrix element given by
\begin{eqnarray}
M(\alpha_n J_n) & = & C\sum_j \theta_j U(J_n j J \frac{1}{2},\frac{3}{2} 1)
\nonumber \\
 & \times &  \langle (p\Xi^-)0p_j 0s_{1/2}|V(p\Xi^-\to \lala)|
(\lala)(0s0p)L\!=\!1\ S\!=\!0\ T\!=\!0\rangle
\label{eq:spme}
\end{eqnarray}
where $\theta_j$ is the spectroscopic amplitude for removal of a $0p$
nucleon with angular momentum $j$ from the $^{11}$B ground state
\cite{ck67}, C is an isospin Clebsch-Gordan coefficient taking the
value $-\sqrt{2/3}$ (for the ground state and first three $p$-shell
$2^+$ states of $^{10}$Be, $\sum_{nj}C^2 \theta_j^2 =2.71$ out of 3 $p$-shell
protons) and U is a unitary recoupling coefficient, with arguments
$(J_{\rm Be}j_p J j_\Xi , J_{\rm B} J_{\lala})$, which ensures that
$j_p$ and $j_\Xi$ are coupled to $J_{\lala}$. The two-body matrix element
is equal to $\sqrt{(2j+1)/12} I_0$. Both the $0p_{3/2}$ and $0p_{1/2}$
amplitudes are important and interfere constructively for $2^+_2$.
The mixing amplitudes $b(\alpha_n J_n)$ and the energy
denominators $\Delta E_p$ are given in Table \ref{tab:gaml}. The total
intensity of $s_\Lambda p_\Lambda$ admixtures associated with the four
$^{10}$Be core states considered is $\sim 0.88$\%.

  The spectroscopic amplitudes for $p$-wave $\Lambda$ emission to final
states of the form $^{10}$Be$(J_n^\pi)\otimes 0s_\Lambda$ are
obtained by recoupling the wave function. After a sum over the
final-state spins is performed, the spectroscopic factors are simply
equal to $b^2(J_n^\pi)$ . Table \ref{tab:gaml} gives the decay energy
$E_\Lambda$ for each core state, the single-particle $\Lambda$ width
for this decay energy, and the partial width $\Gamma_\Lambda$ for
each core state. The total width for $p$-wave $\Lambda$ emission
sums to 398 keV.

  The spectroscopic amplitudes for $p$-wave neutron emission to final
states of the form $^{9}$Be$(J_c)\otimes s_\Lambda p_\Lambda$ are
obtained by removing a neutron from the $^{10}$Be core state and
recoupling the wave function. After a sum over the
final-state spins is performed, the recoupling coefficients disappear
and we are left with
\begin{equation}
 S(0p) = \sum_{\alpha_n\alpha'_n J_n j_p} b(\alpha_n J_n)
b(\alpha_n' J_n) \langle \alpha_n J_n||a_{j_p}^\dagger ||J_c\rangle
\langle \alpha'_n J_n||a_{j_p}^\dagger ||J_c\rangle
\label{eq:s0p}
\end{equation}
Table \ref{tab:gamn} lists the $^9$Be core states of importance, their
excitation energies, the decay energy $E_n$ for each core state, $S(0p)$,
the single-particle neutron width for the decay energy, and the
partial width $\Gamma_n$ for each core state. The total width for
$p$-wave neutron emission sums to 622 keV. The effect of coherence
amongst the admixed amplitudes for different $^{10}$Be $2^+$ states
in Eq.~\ref{eq:s0p} is significant. The summed spectroscopic
strength for neutron removal from each of the first three $^{10}$Be states
to the $^9$Be states listed in Table \ref{tab:gamn} is in the range
$2.5-3.0$ (out of four $p$-shell neutrons), while the fourth $^{10}$Be state
contributes little (these results follow from the supermultiplet
structure of the $p$-shell wave functions).

\subsubsection{$0s_\Lambda^2$ admixtures into the $0s$ $\Xi^-$ state}
\label{ssmix}

 If we took an $s^3p^7$ shell-model configuration, with
$J^\pi ;T$=$1^-;1$ and the $p^7$ configuration being the $p$-shell $3/2^-$
ground state for $^{11}$B, to represent the $0s$ proton hole state
in $^{11}$B, the mixing matrix element in question would be
\begin{eqnarray}
\langle s^4 p^{n-1} s_\Xi |V(p\Xi^-\to \Lambda\Lambda)| s^3 p^{n-1} s_\Lambda^2
\rangle  =  \langle s^4 s_\Xi |V| s^3 s_\Lambda^2 \rangle \nonumber \\
 =  \sqrt{1/2} \langle (p\Xi^-)(0s0s)|V|(\Lambda\Lambda)(0s0s)L\!=\!0\
S\!=\!0\ T\!=\!0\rangle
\label{eq:mes}
\end{eqnarray}
since the $p$-shell configurations factor out of the matrix element.
The matrix element in Eq.~\ref{eq:mes} must be reduced by a center
of mass correction factor which is essentially, but not quite, the
$\alpha$ in Eq.~\ref{eq:ximix}. To make this point explicit, and to
make an analytic but realistic calculation, we take the $^{11}$B
ground state wave function to be the $3/2^-$ state with [443] spatial
symmetry and $K\!=\!3/2$; the SU(3) symmetry is (1\ 3). Namely,
\begin{equation}
|^{11}{\rm B}_{g.s.}\rangle =\sqrt{\frac{21}{26}}|(1\ 3)L\!=\!1
\ S\!=\!\frac{1}{2}\rangle - \sqrt{\frac{5}{26}}|(1\ 3)L\!=\!2
\ S\!=\!\frac{1}{2}\rangle .
\label{eq:gswf}
\end{equation}
This configuration accounts for somewhat over 80\% of the $p$-shell
wave function with a typical effective interaction. Then, the $s^3p^7$
$0s$-hole state with $J^\pi$=$1^-$ and $T$=1 is
\begin{equation}
|s^3\otimes ^{11}{\rm B}_{g.s.}\rangle =\sqrt{\frac{14}{26}}|(1\ 3)L\!=\!1
\ S\!=\!0\rangle - \sqrt{\frac{7}{26}}|(1\ 3)L\!=\!1\ S\!=\!1\rangle +
\sqrt{\frac{5}{26}}|(1\ 3)L\!=\!2\ S\!=\!1\rangle .
\label{eq:shwf}
\end{equation}
The structure of the spurious center of mass states to which the state
in Eq.~\ref{eq:shwf} must be orthogonalized are independent of $L$
and can be written
\begin{eqnarray}
|(1\ 3)T\!=\!1\ S\!=\!0\rangle =-\sqrt{\frac{3}{10}}|s^3p^7\rangle +
\sqrt{\frac{8}{15}}|(3\ 1)\frac{1}{2}\frac{1}{2}\rangle -
\sqrt{\frac{1}{30}}|(1\ 2)\frac{1}{2}\frac{1}{2}\rangle +
\sqrt{\frac{2}{15}}|(1\ 2)\frac{3}{2}\frac{1}{2}\rangle
\nonumber \\
|(1\ 3)T\!=\!1\ S\!=\!1\rangle =\sqrt{\frac{1}{10}}|s^3p^7\rangle +
\sqrt{\frac{1}{10}}|(1\ 2)\frac{1}{2}\frac{1}{2}\rangle +
\sqrt{\frac{4}{10}}|(1\ 2)\frac{1}{2}\frac{3}{2}\rangle +
\sqrt{\frac{4}{10}}|(1\ 2)\frac{3}{2}\frac{1}{2}\rangle ,
\nonumber \\
\label{eq:spur}
\end{eqnarray}
where we have labelled the $s^4p^5(sd)$ configurations by giving
only $(\lambda\ \mu)TS$ for the $p^5$ configuration (the
configurations with $S$=0 and $S$=1 have spatial symmetries [442]
and [433] respectively). The overlaps of the $s^3p^7$ state in
Eq.~\ref{eq:shwf} with the spurious states
of the appropriate $L$ can be read off from Eqs.~\ref{eq:spur}
and the non-spurious $0s$-hole state obtained by Schmidt
 orthogonalization.
The $J^\pi = 2^-$ $0s$-hole state can be similarly constructed and
the total $0s$ occupancy of 1.7 protons in the wave function of
Eq.~\ref{eq:gswf} is shared between the $1^-$ and $2^-$ states in
the ratio $\sqrt{3/5}.\sqrt{103/115}$.

 The overlap between the wave function of Eq.~\ref{eq:shwf} and
the corresponding non-spurious state is $\sqrt{103/130}$ and this
multiplied by $\sqrt{1/2}I_0$ (the two-body matrix element in
Eq.~\ref{eq:mes} is equal to $I_0$) gives the required mixing matrix
element; the mixing matrix element for $I_0 = 2.39$ MeV is
1.507 MeV. Taking $\Delta E_s$ equal to 9.8 MeV from
Fig.~1, we obtain a perturbative estimate of a 2.4\% admixture of the
$^{10}{\rm Be}(s^{-1}_N)\otimes s^2_\Lambda$ configuration ($c^2 =0.024$ in
Eq.~\ref{eq:ximix}).

 The $s_\Lambda^2$ admixtures in the $0s$ $\Xi$ hypernuclear state
provide a mechanism for nucleon decay to $s^4p^5\otimes s_\Lambda^2$
hypernuclear states, in particular for neutron emission which has the
lowest threshold and largest decay energy of 22.3 MeV (9.9 MeV for
proton emission). To estimate, for example, the spectroscopic factors
for neutron decay to $\lala$ hypernuclear states with the $^9$Be core
states listed in Table \ref{tab:gamn}, one looks for the $s^4p^5(sd)$
component in the non-spurious $1^-$ $0s$-hole state which has [41]
spatial symmetry, or $(3\ 1)$ SU(3) symmetry, for the $p^5$ part.
The amplitude for this component, which has $L$=1 and $S$=0,
is $\sqrt{56/515}$. It remains to transform to a $p^5;J_1 T_1\otimes
(sd);j$ representation using
\begin{eqnarray}
|p^5(\lambda_1\ \mu_1)T_1 S_1\otimes (sd)(2\ 0)\frac{1}{2}\frac{1}{2}
;(1\ 3)LSJT\rangle =\sum_{L_1 J_1 lj} \langle(\lambda_1\ \mu_1)
L_1(2\ 0)l||(1\ 3)L\rangle \nonumber \\
\left( \begin{array}{ccc} L_1 & S_1 & J_1 \\
 l   & \frac{1}{2} & j \\
 L   & S   & J
\end{array} \right) |p^5(L_1S_1)J_1T_1\otimes(sd)(l\frac{1}{2})j\rangle.
\label{eq:tran}
\end{eqnarray}
Making an assumption of pure SU(3) LS wave functions for the $^9$Be
states in Table \ref{tab:gamn}, we obtain 131 keV for the neutron width
to these states from the $s_\Lambda^2$ admixture. When the full $p$-shell
wave functions for all energetically available states are used,
the calculated neutron width increases to 155 keV ($\Gamma_n(s) =
61$ keV, $\Gamma_n(d) =93$ keV) and a proton width of 19 keV is
obtained for a total nucleonic decay width of 174 keV.

  The escape width for the $1^-$ $0s$-hole state of $^{10}$Be,
calculated using the decay energies appropriate to $\Xi^-$
hypernuclear decay, is 7.4 MeV ($\Gamma_n =6.6$ MeV). In the
purely nuclear case of a proton-hole state in
$^{11}$B, the maximum neutron energy is estimated to be 15.2 MeV
compared with 22.3 MeV for the hypernuclear case and the calculated
escape width will be correspondingly less; the neutron width for the
$2^-$ state, which would be needed in any comparison with data and
which enters into the calculation of widths for $0p_\Xi$ states,
is about a factor of three smaller than that for the $1^-$ state
because the amplitude of the $p^5[41](sd)$ component in the
wave function, which gives access to the low-lying final states
of $^9$Be, is much smaller.

\subsubsection{Discussion of $0s$ $\Xi^-$ Widths}
\label{swidth}

 The partial decay widths that we calculate for the $1^-$,
$^{11}$B$(g.s.)\otimes 0s_\Xi$ state are collected in Table \ref{tab:wids}
and compared with the corresponding results of Ikeda \etal\
\cite{ikeda94}. Ikeda \etal\ define partial conversion widths
$\Gamma_{bb}$, $\Gamma_{bc}$ and $\Gamma_{cc}$, where $b$ and
$c$ refer to bound and continuum $\Lambda$ states, and present
the contributions from each set of $N$ and $\Lambda$
orbits for the $0s$ and $0p$ $\Xi^-$ states. Their expression
for $\Gamma_{bb}$ is
\begin{eqnarray}
 \Gamma_{bb} = \sum_{L} {{(2L+1)}\over {4(2l_\Xi +1)}}\ {\cal P}(n_N l_N)
\ \langle n_\Xi l_\Xi n_N l_N|v_{\Xi N,\lala}|n_{\Lambda_1} l_{\Lambda_1}
 n_{\Lambda_2} l_{\Lambda_2}\rangle^2_{L\ S\!=\!0\ T\!=\!0} \nonumber \\
{\Gamma_{n_N l_N}^{(h)}\over {(\epsilon_{n_\Xi l_\Xi}+\epsilon_{n_N l_N}
-\epsilon_{n_{\Lambda_1} l_{\Lambda_1}}-\epsilon_{n_{\Lambda_2} l_{\Lambda_2}}
+\Delta)^2 + (\Gamma_{n_N l_N}^{(h)})^2/4}} \label{eq:gambb}
\end{eqnarray}
where $\Delta$ is the $\Xi^- p-\lala$ mass difference,
$\Gamma_{n_N l_N}^{(h)}$ denotes the width of a hole state, and
${\cal P}(n_N l_N)$ is a nucleon occupation probability taking into
account the fact that the core is not an $LS$ closed shell. Similar
expressions for $\Gamma_{bc}$ and $\Gamma_{cc}$ involve integrals over
the momenta of the emitted $\Lambda$ particles.

 We note that the result $\Gamma_{bb}$=551 keV in Table \ref{tab:wids}
is arrived at by endowing the nuclear $0s$-hole state with a width of
10 MeV, taken from the observed width in $^{12}$C. The two-body matrix
element, equal to $\sqrt{2} I_0$ for $N\Xi$ states with good isospin,
has the value 3.45 MeV and the difference of single-particle energies
in the denominator comes to 5.39 MeV. Our expression can be written in
a form similar to that of Eq.~\ref{eq:gambb} with the width term removed
from the Breit-Wigner denominator, the full $B_{\lala}$ of
Eq.~\ref{eq:bll} in place of the $\Lambda$ single-particle energies
and the escape width for the $1^-$ $0s$-hole state of $^{10}$Be,
calculated using the decay energies appropriate to $\Xi^-$
hypernuclear decay, in the numerator. The neutron escape width
is calculated to be 6.6 MeV ($\Gamma_n +\Gamma_p = 7.4$ MeV) which,
when multiplied by the admixed intensity of 0.02366, gives the neutron
width of 155 keV listed in Table \ref{tab:wids}. The main difference between
the two results lies in the additional 4.5 MeV from $\Delta B_{\lala}$
in the energy denominator. The remaining difference is due to the center
of mass factor (0.79), the difference in widths for the $0s$-hole state
and a small difference in two-body matrix elements.

 Our $p$-wave $\Lambda$ widths are in modest agreement with those of Ikeda
\etal\ In our treatment, the $0s$ $\Lambda$ has a parentage to escape
leaving a $\Lambda$ hypernucleus with the $\Lambda$ in a $0p$ state.
We label this channel by $p_N^{-1} p_\Lambda^b s_\Lambda^c$ in
Table \ref{tab:wids} and the continuum channels treated by Ikeda
\etal\ by $p_N^{-1} p_\Lambda^b (1s0d)_\Lambda^c$.
For the latter channels, we give no entry for our calculation since
we do not wish to treat the
unbound $(sd)_\Lambda$ orbits using harmonic oscillator wave
functions. In this approximation, we would have the additional difficulty
that the energies of some of these states would be essentially degenerate
with the $\Xi^-$ state. Finally, when we treat the nuclear $0s$-hole
state as a discrete state, $\Lambda$ decay channels are closed on account
of the high excitation energy of the residual $\Lambda$ hypernucleus.

 By far the biggest difference between the two calculations arises
for $p$-wave neutron decay. Our relatively large partial width
for this channel can be understood in comparison to the calculated
width for $p$-wave $\Lambda$ emission. Both widths arise from
$s_\Lambda p_\Lambda$ admixtures in the $\Xi^-$ state and roughly two
of the four $p$-shell neutrons are linked to low-lying states of
$^9$Be in the parentage decomposition of the $^{10}$Be core states.
In comparison, there is only a single $p$-shell $\Lambda$ to be emitted.
For this rough 2:1 ratio in spectroscopic strength for neutron versus
$\Lambda$ decay, a very good account of the relative $\Gamma_n$
and $\Gamma_\Lambda$ values in Table \ref{tab:wids} is obtained once the
differences in $\Gamma_{sp}$ from Tables \ref{tab:gaml} and \ref{tab:gamn}
for the somewhat different decay energies are taken into account.
On the other hand, Ikeda \etal\ endow $0p$-hole states with a width
which is based on the distribution of pickup strength from $^{11}$B
and use Eq.~\ref{eq:gambb}. This does not seem correct since the
$^{10}$Be core states of interest are discrete states, which can be
treated as such in bound-state shell-model calculations, and their
spacings have nothing to do with decay widths from the $\Xi$ state
beyond their influence in the energy denominators for perturbative
estimates of $s_\Lambda p_\Lambda$ admixtures. In fact, in the limit
in which the entire $0p$-hole strength is concentrated in one state
of the residual nucleus, namely when $\Gamma^{(h)}_{op}\to 0$,
expression (\ref{eq:gambb}) leads to the unacceptable result of a zero
width.

\subsubsection{Widths of $0p$ $\Xi^-$ states}
\label{pwid}

  The $2^+$ states that are expected to be most strongly formed
in the $^{12}$C$(K^-,K^+)$ reaction have a $p_{3/2}$ or $p_{1/2}$
$\Xi^-$ coupled to the $^{11}$B ground state. In analogy to
$^{12}_{\ \Lambda}$C, the two configurations are expected to mix
to form two $2^+$ states with a separation of the order of an MeV
and a ratio of cross sections which depends on just how strongly
the basis configurations, which have equal strength, are mixed.

  In shell-model terms, the $0p$ $\Xi$ states are $2\hbar\omega$
states with respect to the $\lala$ hypernuclear ground state.
This means that the $0p$ $\Xi$ state can mix with the $0\hbar\omega$
$s_{\Lambda}^2$ states or with $2\hbar\omega$ $\lala$ states. In the
$0\hbar\omega$ case, the energy denominators are very large,
e.g. 41 MeV for $^{10}$Be$(gs)\otimes s_\Lambda^2$, and the
admixtures will be small. In the $2\hbar\omega$ case, there are many
different types of configurations according to the way in which quanta
are distributed amongst the nuclear and $\Lambda$ orbitals. All the
different types of configurations are listed in Table \ref{tab:conf},
together with the kinds of two-body matrix elements
which admix them into the $p^7p_{\Xi}$ configuration of interest.
Expressions for the two-body matrix elements are given in
Table \ref{tab:talmi}.

 Many features of the decay scheme for the $p_\Xi$ states will be
similar to that for the $s_\Xi$ state since the increased excitation
energy, $\epsilon_p -\epsilon_s$ for the $\Xi$, is balanced, for the
most important admixed configuration ($s^4p^6p_\Lambda^2$), by having
an extra $p_\Lambda$ in the admixed state and the final A-1 system.
Given the rather similar spacings of the $s_\Xi$,$p_\Xi$ and
$s_\Lambda$,$p_\Lambda$ orbits, a very crude estimate would give
\begin{eqnarray}
\Gamma_{p_\Lambda}(p_\Xi) \sim {\cal R}^2 \ 2 \Gamma_{p_\Lambda}(s_\Xi)
\label{eq:plwid} \\
\Gamma_{p_N}(p_\Xi) \sim {\cal R}^2 \  \Gamma_{p_N}(s_\Xi)\ , \label{eq:pnwid}
\end{eqnarray}
where ${\cal R}$ is the ratio of mixing matrix elements induced by the
$\langle p_Np_\Xi|V|p_{\Lambda}^2\rangle$ and
$\langle p_Ns_\Xi|V|s_{\Lambda}p_\Lambda\rangle$ interactions. The
factor of 2 in Eq.~\ref{eq:plwid} appears because either $p_\Lambda$
from the $p_\Lambda^2$ configuration can escape.

 The essential features of Eqs.~\ref{eq:plwid} and \ref{eq:pnwid} are
apparent from an examination of our calculated partial widths, which are
presented in Table \ref{tab:widp}. Results are presented for both
$0p$ $\Xi^-$ basis states, namely a $p_{3/2}$ or a $p_{1/2}$ cascade
coupled to the $^{11}$B ground state. If the two basis states are
allowed to mix, the summed partial widths for the $p_{3/2}$ and $p_{1/2}$
states will be shared by the two resultant $2^+$ states. The differences
in the partial widths for the two basis states can be understood in terms
of the values for the two-body matrix elements in Table \ref{tab:talmi}
and some peculiarities of the recoupling coefficient analogous to the
one in Eq.~\ref{eq:spme}. First consider the $\Gamma_\Lambda$ widths
which result from admixing $p_N^{-1}s_\Lambda (sd)_\Lambda$ configurations,
bearing in mind our reservation about treating the $(sd)_\Lambda$
orbits in a bound-state approximation. For the $1s_\Lambda$ orbit,
$L_{\lala}$=0 and the mixing matrix element is very small (Table
\ref{tab:talmi}). For the $d_\Lambda$ orbit, $L_{\lala}$=2 and the mixing
matrix element is substantial. However, the recoupling  coefficient vanishes
for a $p_{3/2}$ $\Xi$ when the core is in a $2^+$ state but is large
for a $p_{1/2}$ $\Xi$. The same remarks apply for the $p_N^{-1}
p_\Lambda^2$ configuration except that now there is a large two-body
matrix element for $L_{\lala}$=0 which acts to enhance $\Gamma_\Lambda$
and $\Gamma_n$ for a $p_{3/2}$ $\Xi$.

  The total decay widths obtained for the $p_{3/2}$ and $p_{1/2}$ cascade
states are of the order 1.8 MeV and 1.6 MeV, respectively, which are
somewhat larger than the width of 1.2 MeV obtained for the $0s$
$\Xi$ state. These widths should be increased slightly to take into
account the effect of unbound $\Lambda$ orbits that we do not
consider in our essentially bound-state approach. The total width
associated with the pair of $2^+$ states will be larger than that
for either of the individual states.

 Once again, Ikeda \etal\ obtain very small neutron widths. Also, it
is puzzling that they obtain such a small $\Lambda$ width from the
$p_N^{-1} p_\Lambda^b p_\Lambda^c$ configuration given the
substantial width that they obtain for the $0s$ $\Xi$ state from
the $p_N^{-1} s_\Lambda^b p_\Lambda^c$ configuration. The end result
is that they obtain a width for the $0p$ $\Xi$ state which is
considerably smaller than that for the $0s$ $\Xi$ state. In addition,
it is not clear which state their calculated $0p_\Xi$ width
applies to.

\section{Summary}
\label{summary}

  The purpose of the present paper was to obtain an estimate for the
widths of $0s$ and $0p$ $\Xi^-$ states which can be produced in the
$^{12}$C$(K^-,K^+)$ reaction. We used standard shell-model techniques
to make estimates of the mixing of $\Xi$ and $\lala$ hypernuclear states
and hence obtain estimates for the neutron and $\Lambda$ decay
widths of the $\Xi$ hypernuclear states. For a $\Xi^-p\to\lala$
interaction based on G matrices calculated using a version of the
Nijmegen Model D interaction for the baryon-baryon interaction
in the strangeness -2 sector, decay widths in the range $1-2$ MeV
were obtained for these states.

  A large part of the decay width for the $\Xi$ hypernuclear states is
attributed to $p$-wave emission of both neutrons and $\Lambda$ particles
following the conversion of the $\Xi^-$ and a $p$-shell proton. The
nuclear core states on which the admixed $\lala$ hypernuclear states
are based are essentially discrete states which are directly amenable
to our straightforward shell-model treatment. On the other hand,
the nuclear $0s$-hole strength, which is relevant when the $\Xi^-$
converts on an $s$-shell proton, appears as a broad resonance-like
distribution in knockout reactions. In realistic shell-model calculations,
the $0s$ hole strength is spread over a considerable range of excitation
energy with each fragment having a substantial decay width into
nucleon, and perhaps other, channels. For our estimates of $\Xi$
hypernuclear decay widths, we have concentrated the $0s$-hole strength
in a single state related to the hypernuclear core state and have, in effect,
calculated the neutron decay width of this state for the energetics
determined by the energy of the $\Xi$ hypernuclear state. This should give
quite a reasonable estimate of the escape width for the $\Xi$ state.
If the $\Xi$ state comes close in energy to $\lala$ states based on
major fragments of the $0s$-hole strength, it could be fragmented
by configuration mixing and thus acquire a spreading width. This point
needs to be investigated in more detailed shell-model calculations.
However, the separation of about 10 MeV between the $\Xi$ states of
interest for A=12 and the centroid of $\lala$ states based on nuclear
$0s$-hole states suggests that the $\Xi$ states will not be appreciably
fragmented by this mechanism. For somewhat heavier systems, this may not
be the case as the $0s$-hole state becomes more deeply bound. For example,
we estimate that the 10 MeV separation for A=12 is reduced to about 3 MeV
for A=16.

 A caveat to the results obtained in this paper is that the decay widths
of $\Xi$ hypernuclear states depend
quadratically on the strength of the effective $\Xi N-\lala$ interaction
in the medium, about which there is, at present, considerable uncertainty.
If the strength is similar to that deduced from G-matrix
calculations \cite{him93} using the Nijmegen Model D interaction,
there should exist $\Xi$ hypernuclear states narrow enough for study
using the $(K^-,K^+)$ reaction with a resolution of $\sim 2$ MeV. Of
course, it must also be remembered that we have assumed the existence
of a $\Xi$-nucleus potential deep enough to bind $\Xi$ single-particle
states.
For in-flight studies of the $(K^-,K^+)$ reaction, the high-quality and
intensity of $K^-$ beams from the D6 beamline \cite{pile92} at the
Brookhaven AGS could provide a promising means to search for
$\Xi$ hypernuclear states. However, it is evident from the discussion
in Sec.~\ref{singlepart} that such studies are only on the borderline
of feasibility and an improved spectrometer with high efficiency
and good resolution will be required. As emphasised in the first two
sections of this paper, the physics to be gained from a successful
observation of $\Xi$ hypernuclear states is considerable.

\section*{Acknowledgement}

This work was supported by the U.S. Department of Energy under Contract
DE-AC02-76CH00016. One of us (AG) was partially supported
by an award from the Alexander von Humboldt Foundation and by the Basic
Research Foundation of the Israeli Academy of Sciences and Humanities.


\pagebreak
\setlength{\parindent}{0.0cm}
\begin{center}
{\Large \bf Tables}
\end{center}

\begin{table}[h]
\caption[a]{$s_\Lambda p_\Lambda$ mixing amplitudes ($b$) and $p$-wave
partial $\Lambda$ decay widths $(\Gamma_\Lambda)$for the $1^-$ $0s$
$\Xi^-$ state of $^{12}_\Xi$Be.  $\Gamma_{sp}$ for a decay energy $E_\Lambda$
is calculated for an equivalent square well of radius $R = 3.33$ fm, depth
$V_0 = 25 $ MeV and reflection factor $f = 2.5$ (see Eq.~\ref{eq:spw}).
All energies are in MeV. The total admixed intensity is 0.88\% and
the total $p$-wave $\Lambda$ decay width is 398 keV.}
\vspace{0.1cm}
\begin{center}
\begin{tabular}{cccccc}
\hline
Core State & $\Delta E_p$ & $b(\alpha_n J_n^\pi)$ & $E_\Lambda$ & $\Gamma_{sp}$
& $\Gamma_{\Lambda}$ \\
\hline
 $0^+_1$ & 21.5 & -0.0422 & 16.8 & 56.4 & 0.100 \\
 $2^+_1$ & 18.1 & -0.0440 & 13.4 & 48.7 & 0.094 \\
 $2^+_2$ & 15.5 & ~0.0654 & 10.8 & 41.9 & 0.179 \\
 $2^+_3$ & 12.1 & -0.0275 & ~7.4 & 31.0 & 0.024 \\
\hline
\end{tabular}
\end{center}
\label{tab:gaml}
\end{table}

\begin{table}[h]
\caption[a]{Neutron spectroscopic factors ($S$) and $p$-wave
partial neutron decay widths $(\Gamma_n)$for the $1^-$ $0s$
$\Xi^-$ state of $^{12}_\Xi$Be.  $\Gamma_{sp}$ for a decay energy $E_n$
is calculated for an equivalent square well of radius $R = 3.33$ fm, depth
$V_0 = 40 $ MeV and reflection factor $f = 2.5$ (see Eq.~\ref{eq:spw}).
All energies are in MeV. The total $p$-wave neutron decay width is 622 keV.}
\vspace{0.1cm}
\begin{center}
\begin{tabular}{cccccc}
\hline
$^9$Be State & $E_x(^9$Be) & $S$ & $E_n$ & $\Gamma_{sp}$
& $\Gamma_n$ \\
\hline
 $\frac{3}{2}^-_1$ & 0.0 & 0.00617 & 12.0 & 43.6 & 0.269 \\
 $\frac{5}{2}^-_1$ & 2.4 & 0.00054 & ~9.6 & 36.9 & 0.020 \\
 $\frac{1}{2}^-_1$ & 2.8 & 0.00168 & ~9.2 & 35.7 & 0.060 \\
 $\frac{3}{2}^-_2$ & 4.7 & 0.00047 & ~7.3 & 29.5 & 0.014 \\
 $\frac{7}{2}^-_1$ & 6.4 & 0.00806 & ~5.6 & 23.2 & 0.187 \\
 $\frac{5}{2}^-_2$ & 7.4 & 0.00379 & ~4.6 & 19.0 & 0.072 \\
\hline
\end{tabular}
\end{center}
\label{tab:gamn}
\end{table}
\pagebreak

\begin{table}[h]
\caption[a]{Summary of partial contributions to the width of a
$J^\pi$=$1^-$, $0s_\Xi$ state in $^{12}_{\Xi^-}$Be in the calculations
of Ikeda {\it et al.} \cite{ikeda94} and this work. The different
contributions are labelled in the manner of Ref.~\cite{ikeda94}
by $l_N^{-1}l_{\Lambda}^i l_{\Lambda}^j$ with $i$,$j$=$b$,$c$ where
$b$ and $c$ mean a $\Lambda$ particle in bound and continuum states,
respectively. We compare the $\Gamma_{bb}$ and $\Gamma_{bc}$ of
Ref.~\cite{ikeda94} with our values for $\Gamma_n\simeq \Gamma_N$
and $\Gamma_\Lambda$. In addition, we give the maximum decay energies
$E_n$ and $E_\Lambda$, using our kinematics, for each partial
decay channel. With our use of a discrete energy for the nuclear
$0s$-hole state, some channels are closed. An asterisk indicates
channels with unbound $\Lambda$ orbits that we do not treat.}
\vspace{0.1cm}
\begin{center}
\begin{tabular}{cccccccc}
\hline
Channel & $\Gamma_{bb}$ & $E_n$ & $\Gamma_n$ & Channel &
$\Gamma_{bc}$ & $E_\Lambda$ & $\Gamma_\Lambda$ \\
 & keV & MeV & keV & & keV & MeV & keV \\
\hline
 $s_N^{-1}s_\Lambda^b s_\Lambda^b$ & 551 & 22.3 & 155 & $s_N^{-1}
 s_\Lambda^b s_\Lambda^c$ & 111 & $<$0 & \\
 $s_N^{-1}p_\Lambda^b p_\Lambda^b$ & ~17 & ~1.7 & & $s_N^{-1}
 p_\Lambda^b p_\Lambda^c$ & ~30 & $<$0 & \\
 $p_N^{-1}s_\Lambda^b p_\Lambda^b$ & ~25 & 12.0 & 622 & $p_N^{-1}
 s_\Lambda^b p_\Lambda^c$ & 281 & 16.8 & 398 \\
 & & & & $p_N^{-1}p_\Lambda^b s_\Lambda^c$ & & 6.5 & 158 \\
 & & & & $p_N^{-1}p_\Lambda^b 1s_\Lambda^c$ & ~23 & 6.5 & $*$ \\
 & & & & $p_N^{-1}p_\Lambda^b d_\Lambda^c$ & 118 & 6.5 & $*$ \\
\hline
\end{tabular}
\end{center}
\label{tab:wids}
\end{table}

\begin{table}[h]
\caption[a]{Configurations and two-body mixing matrix elements which
enter into a shell-model calculation for studies of the $0p$ $\Xi^-$
states.}
\vspace{0.1cm}
\begin{center}
\begin{tabular}{ll}
\hline
 Configurations & Two-body matrix elements \\
\hline
 $s^4p^7p_\Xi$    &   \\
$(\alpha s^3p^8+\beta s^4p^6(sd))s_\Xi$ & $\langle s_Np_\Xi|V|p_Ns_\Xi
\rangle$,\ $\langle p_Np_\Xi|V|(sd)_Ns_\Xi\rangle$ \\
 $s^4p^6p_{\Lambda}^2$ & $\langle p_Np_\Xi|V|p_{\Lambda}^2\rangle$ \\
 $s^4p^6s_{\Lambda}(sd)_\Lambda$ & $\langle p_Np_\Xi|V|
s_{\Lambda}(sd)_\Lambda\rangle$ \\
$(\alpha' s^3p^7+\beta' s^4p^5(sd))s_{\Lambda}p_\Lambda$ &
 $\langle s_Np_\Xi|V|s_{\Lambda}p_\Lambda\rangle$ \\
$(\alpha'' s^4p^4(sd)^2+\beta'' s^3p^6(sd)+\gamma'' s^4p^5(pf))s_\Lambda^2$
 & 0 \\
 $s^4p^6s_{\Lambda}^2$ & $\langle p_Np_\Xi|V|s_{\Lambda}^2\rangle$ \\
\hline
\end{tabular}
\end{center}
\label{tab:conf}
\end{table}
\pagebreak

\begin{table}[h]
\caption[a]{Expessions in terms of Talmi integrals for two-body mixing
 matrix elements which
enter into a shell-model calculation for studies of the $0p$ $\Xi^-$
states. The $N\Xi$ configurations are proton-$\Xi^-$ in nature.
They and the $\lala$ configurations have $S$=0 and the $L$ specified
in column 2. The values of the two-body matrix elements for the
interaction that we use are given in the final column.}
\vspace{0.1cm}
\begin{center}
\begin{tabular}{lclc}
\hline
Two-body matrix elements & $L$ & Expression & Value (MeV) \\
\hline
$\langle p_Np_\Xi|V|p_{\Lambda}^2\rangle$ & 0 & $\frac{1}{4}(5I_0 -6I_1
+5I_2)$ & ~2.607  \\
$\langle p_Np_\Xi|V|p_{\Lambda}^2\rangle$ & 2 & $\frac{1}{2}(I_0 +I_2)$
 & ~1.215  \\
 $\langle p_Np_\Xi|V|0s_{\Lambda}1s_\Lambda\rangle$ & 0 & $-\frac{1}{4}
(I_0 -6I_1 +5I_2)$ & $-0.212$ \\
 $\langle p_Np_\Xi|V|0s_{\Lambda}0d_\Lambda\rangle$ & 2 & $\frac{1}{2}
(I_0 -I_2)$  & ~1.180 \\
 $\langle s_Np_\Xi|V|s_{\Lambda}p_\Lambda\rangle$ & 1 & $\sqrt{\frac{1}{2}}
I_0$  & ~1.693 \\
 $\langle p_Np_\Xi|V|s_{\Lambda}^2\rangle$ & 0 & $-{\sqrt{3}\over 2}
(I_0 -I_1)$  & $-1.826$ \\
\hline
\end{tabular}
\end{center}
\label{tab:talmi}
\end{table}

\begin{table}[h]
\caption[a]{Summary of partial contributions to the width of a
$J^\pi$=$2^+$, $0p_\Xi$ states in $^{12}_{\Xi^-}$Be in the calculations
of Ikeda {\it et al.} \cite{ikeda94} and this work, which includes
results for both $0p_{3/2}$ and $0p_{1/2}$ $\Xi^-$ particles
coupled to the $^{11}$B ground state. See the caption to
Table \ref{tab:wids} for notation.}
\vspace{0.1cm}
\begin{center}
\begin{tabular}{cccccccccc}
\hline
Channel & $\Gamma_{bb}$ & $E_n$ & $\Gamma_n$($p_{3/2}$) &
$\Gamma_n$($p_{1/2}$) & Channel & $\Gamma_{bc}$ & $E_\Lambda$ &
$\Gamma_\Lambda$($p_{3/2}$) & $\Gamma_\Lambda$($p_{1/2}$) \\
 & keV & MeV & keV & keV & & keV & MeV & keV & keV \\
\hline
 $s_N^{-1}s_\Lambda^b p_\Lambda^b$ & 104 & 21.2 & ~64 & ~84 &
 $s_N^{-1}s_\Lambda^b p_\Lambda^c$ & 115 & ~4.0 & 268 & 268 \\
 $p_N^{-1}s_\Lambda^b s_\Lambda^b$ & ~~1 & 31.5 & ~89 & ~11 &
 $s_N^{-1}p_\Lambda^b 1s_\Lambda^c$ & ~~1 & $<$0 &  & \\
 $p_N^{-1}p_\Lambda^b p_\Lambda^b$ & ~~7 & 10.9 & 368 & 267 &
 $s_N^{-1}p_\Lambda^b d_\Lambda^c$ & ~~8 & $<$0 &  & \\
 $p_N^{-1}s_\Lambda^b sd_\Lambda^b$ & & 10.9 & ~97 & 205 &
 $p_N^{-1}s_\Lambda^b s_\Lambda^c$ & & 26.0 & ~89 & ~13 \\
 & & & & & $p_N^{-1}s_\Lambda^b 1s_\Lambda^c$ & ~~1 & 26.0 & ~~3 & ~~1 \\
 & & & & & $p_N^{-1}s_\Lambda^b d_\Lambda^c$ & ~69 & 26.0 & ~85 & 284 \\
 & & & & & $p_N^{-1}p_\Lambda^b p_\Lambda^c$ & & 15.7 & 738 & 508 \\
 & & & & & $p_N^{-1}p_\Lambda^b 1p_\Lambda^c$ & ~57 & 15.7 & * & * \\
 & & & & & $p_N^{-1}p_\Lambda^b f_\Lambda^c$ &  ~17 & 15.7 & * & * \\
\hline
\end{tabular}
\end{center}
\label{tab:widp}
\end{table}


\pagebreak
\setlength{\parindent}{0.0cm}
\begin{center}
{\Large \bf Figure captions}
\end{center}

\vspace{0.4cm}

{Fig.~1:} {Energy spectrum and decay thresholds of $\Xi$ and $\lala$
hypernuclear configurations for the case of a $^{12}$C target.
The cross hatched area indicates that the nuclear $0s$-hole strength
is fragmented. The $0^+$ and $2^+$ designations at the right of the
figure refer to states of the $^{10}$Be core in $\lala$ hypernuclear
states. The main neutron and $\Lambda$ decay modes of the $0s$
$\Xi^-$-hypernuclear state are indicated in the center of the figure.}

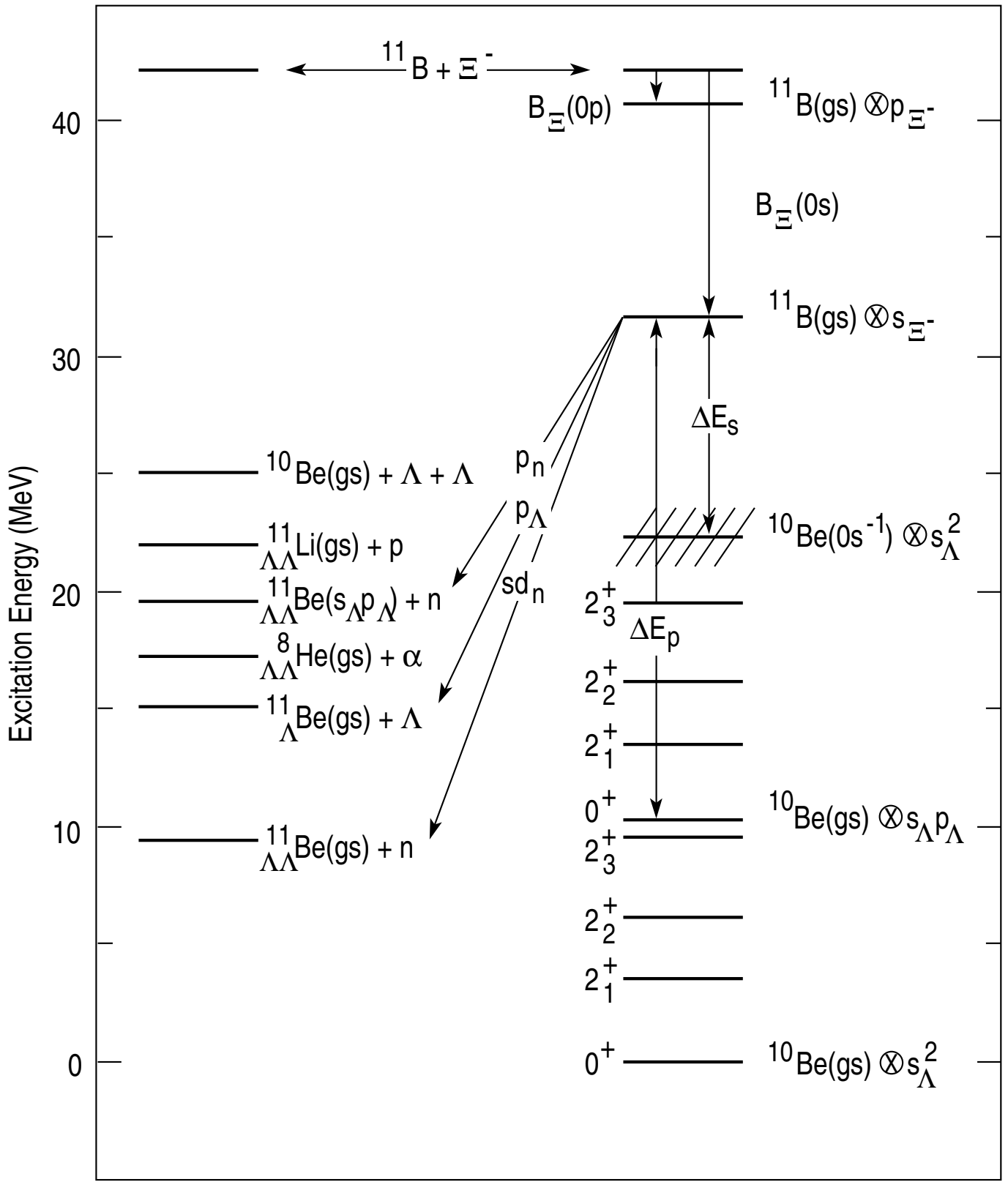
\end{document}